# Optimized Design of Silicon Heterojunction Solar Cells for Field Operating Conditions

Jean Cattin, Olivier Dupré, Brahim Aïssa, Jan Haschke, Christophe Ballif, and Mathieu Boccard

*Abstract*—Solar modules are currently characterized at standard test conditions (STC), defined at 1000 W/m² and 25 °C. However, solar modules in actual outdoor operating conditions typically operate at lower illumination and higher temperature than STC, which significantly affects their performance ratio (average harvesting efficiency over efficiency in STC). Silicon heterojunction (SHJ) technology displays both good temperature coefficient and good low-illumination performances, leading to outstanding performance ratios. We investigate here SHJ solar cells that use a-SiC$_x$(n) layer as front doped layer with different carbon contents under different climates conditions. Adding carbon increases transparency but also resistive losses at room temperature (compared with carbon-free layers), leading to a significant decrease in efficiency at STC. We demonstrate that despite this difference at STC, the difference in energy harvesting efficiency is much smaller in all investigated climates. Furthermore, we show that a relative gain of 0.4%–0.8% in harvesting efficiency is possible by adding a certain content of carbon in the front (n) layer, compared with carbon-free cells optimized for STC.

## I. Introduction

SOLAR cells and modules are generally characterized at—and thus optimized for—standard test conditions (STC), which correspond to a temperature of 25 °C and an illumination of 1000 W/m² with the AM1.5G spectrum. The power output under these test conditions leads to the Watt-peak rating that nowadays dictates the selling price of modules. However, these conditions very rarely correspond to the actual operating conditions of a module in the field, which typically operates at higher temperatures and/or lower illumination than STC [1]–[5]. The new IEC 61853-4 standard on module energy rating includes this perspective by requiring multiple temperature-illumination measurements [6]. This is illustrated in Fig. 1(a)–(e) which shows the yearly produced energy for a typical installation as a function of the operating conditions. The conditions follow from the different climates at the different locations, as shown in Fig. 1(f). These maps are further explained in the discussion, but the following two important observations should be highlighted.

1) STC is not a representative operating condition set in any climate, as, over a year, almost no energy is generated at this particular condition.
2) In different climates, the energy is harvested under different conditions, yet most of the relevant conditions correspond to either higher temperatures than STC or lower irradiance (or both).

Such changes of operating conditions have implications on the solar cell performances. To take field conditions into account, we define the harvesting efficiency ($\eta_\text{harvesting}$) as the yearly produced energy divided by the cumulated inplane irradiance

$$\eta_\text{harvesting} [\%] = \frac{\text{Energy produced [Wh/m}^2\text{]}}{\text{Inplane irradiance [Wh/m}^2\text{]}}. \quad (1)$$

The performance ratio, commonly used to characterize a module in a specific climate, is defined as the harvesting efficiency in a given climate divided by the STC efficiency [7]. All silicon solar cells are subject to a fundamental efficiency decrease upon increasing temperature, at a rate defined as the temperature coefficient (TC) [2], [8]. As shown in [9], the series resistance can also have an impact on the TC of solar modules. This efficiency decrease is mainly driven by the increasing intrinsic carrier concentration, leading to the lower operating voltage that is not balanced by the slightly increasing current. The general approach to maximize the efficiency at high temperature is to maximize the efficiency at STC and minimize TC. One way to achieve the latter is to increase the solar cell's $V_\text{oc}$, as described by the relation proposed by Green *et al.* [10]

$$\text{TC}_{V_\text{oc}} = -\frac{\frac{E_{g0}^\text{Si}}{q} - V_\text{oc}^\text{STC} + \gamma kT}{T}. \quad (2)$$

The requirement for a high $V_\text{oc}$ makes the SHJ technology one of the most interesting candidates for hot climate operation [11], [12], thanks to the excellent surface passivation provided by the thin intrinsic hydrogenated amorphous silicon layers [a-Si(i)] yielding $V_\text{oc}$ values above 730 mV and resulting in TC$_\eta$ in the range of −0.23 to −0.29% K$^{-1}$ [13], [14], the differences being linked in parts to the phenomena described further below. SHJ solar cell technology was already demonstrated to be more

Manuscript received May 17, 2019; revised July 11, 2019; accepted August 27, 2019. This work was supported in part by Qatar Foundation, in part by the European Union's Horizon 2020 research and innovation programs under Grant 745601 (Ampere), and in part by the Swiss National Science Foundation under Ambizione Energy grant ICONS (PZENP2_173627). *(Corresponding author: Jean Cattin.)*

J. Cattin, O. Dupré, J. Haschke, C. Ballif, and M. Boccard are with the Photovoltaics and Thin-Film Electronics Laboratory, Institute of Microengineering, Ecole Polytechnique Fédérale de Lausanne, Neuchâtel 2000, Switzerland (e-mail: jean.cattin@epfl.ch; olivier.dupre@epfl.ch; jan.haschke@epfl.ch; christophe.ballif@epfl.ch; mathieu.boccard@epfl.ch).

B. Aïssa is with the Qatar Environment and Energy Research Institute, Hamad bin Khalifa University, Qatar Foundation, Doha 34110, Qatar (e-mail: baissa@hbku.edu.qa).

Color versions of one or more of the figures in this article are available online at http://ieeexplore.ieee.org.

Digital Object Identifier 10.1109/JPHOTOV.2019.2938449

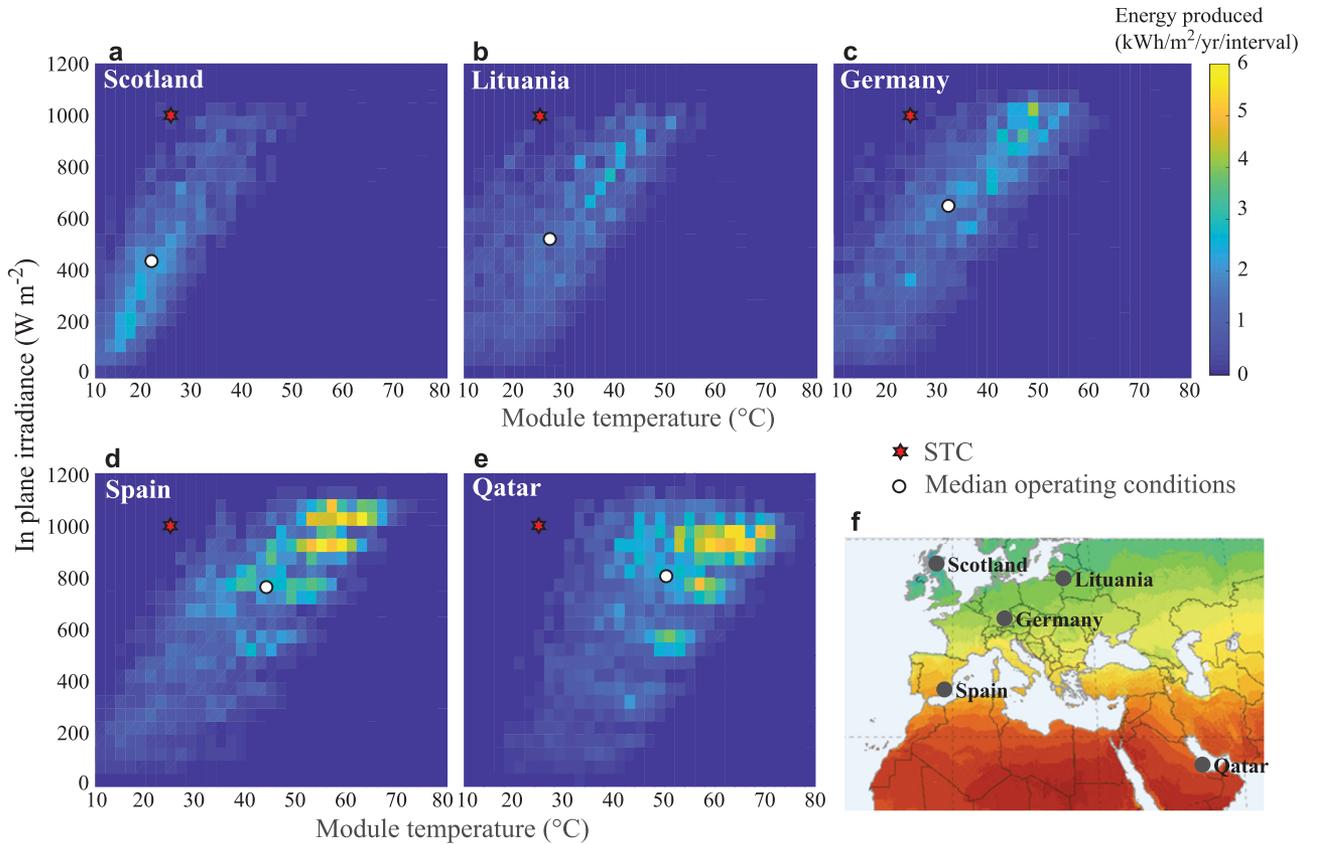

Fig. 1. (a)–(e) Yearly produced energy as a function of the module temperature and illumination in different locations. The STC is indicated with a red star, and the white dot represents the mean operating conditions weighted with the produced energy. (f) Map of the locations used in this article.

efficient in hot climates than equivalent-STC-efficiency modules with other types of silicon-based cells [9].

Another specificity of SHJ cells is the presence of nonOhmic resistances that hinder both the collection of electrons and holes toward their respective electrodes. These resistances are typically attributed to energy barriers in the form of band offsets at the c-Si/a-Si interface and Schottky barrier at the a-Si/transparent conductive oxide interface [15]–[18]. It was shown that the charge transport through these barriers is thermally activated [19]. Additionally, the resistivity of the a-Si layers decreases with increasing temperature [20]. These effects result in a nonlinear FF behavior with respect to the temperature. When cooling the solar cell, FF stops linearly increasing at a certain temperature, eventually leading to an FF maximum and a subsequent decrease at low temperature [19], [21]–[23]. Many SHJ solar cells suffer from such thermally activated $R_s$ already at 25 °C, typically leading to a modest FF below 80% which was for long a ceiling value [24]–[26]. In more severe cases, an S-shaped current–voltage (I–V) can be observed, which highlights that thermally activated $R_S$ components are often nonOhmic [27]–[29]. When present, these thermally activated $R_S$ components usually decrease upon increasing temperature. This leads to an increased FF with higher temperature, which results in a further improved $TC_\eta$ for SHJ compared with standard technologies as discussed in [9] and [30].

Some studies aimed at solar cell architectures leading to a larger current at the expense of a lower FF at 25 °C, with an FF benefitting from thermally activated $R_s$ at higher temperature [17], [30]. Seif et al. added oxygen in the front a-Si(i) layer to obtain a wider bandgap and mitigate parasitic absorption, leading to a current increase [31]. The wider bandgap of the intrinsic layer was, however, accompanied by a large increase of the c-Si/a-SiO valence band offset and larger resistive losses upon hole extraction. The resulting FF loss was shown to be mitigated by increasing the temperature, while the benefit of the larger current was preserved. A higher efficiency for the cell incorporating minute amounts of oxygen in the a-Si(i) layer was thus demonstrated above 50 °C. A similar observation was made in cells using a $MoO_x$ layer as transparent hole selective contact instead of a (p) doped layer [29]. A similar approach was followed by Haschke et al. [32] using rear-junction devices with an n-doped nanocrystalline silicon oxide [nc-SiO(n)] front layer. Similarly, the modified cells benefitted from a current increase (up to 41 mA/cm$^2$ at STC) yet an FF drop. The latter was mitigated at 60 °C, yet not sufficiently to yield an efficiency improvement, and the cells with a nc-SiO(n) layer showed an overall lower efficiency also at 60 °C.

In this article, we go one step further and investigate not only the efficiency at various temperatures under 1-sun illumination, but compare yearly energy harvesting efficiencies in different climates, thus accounting for both temperature and illumination intensity. To highlight the efficiency shift as a function of the temperature and illumination, we use cells with various concentrations of carbon in the front (n)-doped layer. Adding carbon results in a bandgap widening, yielding a higher transparency but hampers charge extraction. Although this is shown to result in an overall performance decrease at STC, the higher operating temperature and/or lower irradiance during operation makes the solar cell incorporating carbon outperform the carbon-free solar cell in terms of yearly harvesting efficiency in all considered climates.

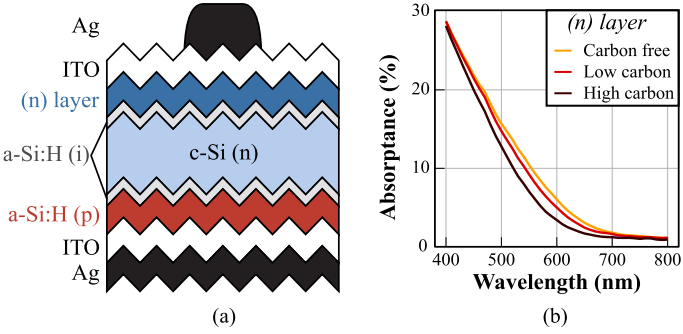

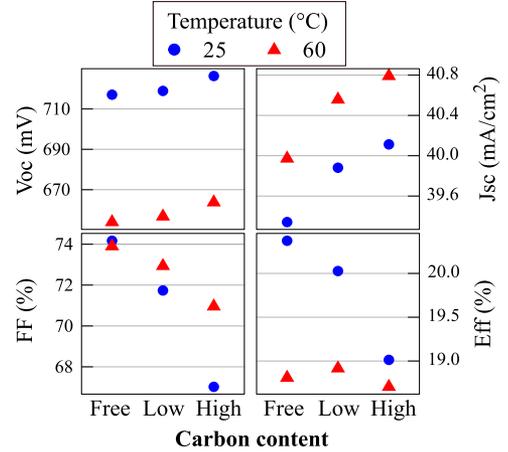

Fig. 2. (a) Scheme of a typical rear-junction SHJ cell. (b) Absorptance of the three investigated (n) doped layers in the visible range.

Fig. 3. I–V characteristics of the cells with various carbon content in the front (n) layer at 25 and 60 °C under 1-sun illumination.

## II. EXPERIMENTAL DETAILS

Fig. 2(a) shows a schematic view of a typical sample. The solar cells presented in this article were all fabricated using float-zone, 180-$\mu$m-thick, n-type (2 $\Omega$cm) silicon wafers, textured by alkaline etching. Before the a-Si layers deposition, an HF solution was used to remove native oxide from the surfaces. The a-Si layers were deposited in a plasma-enhanced chemical vapor deposition reactor. For the carbide n-type layers, $CH_4$ was added to the gas mixture composed of $SiH_4$, $H_2$, and $PH_3$. Methane to silane ratio of 0, 0.4, and 1 were used for the samples presented below, referred, respectively, as carbon-free (reference), low-carbon, and high-carbon.

ITO layers and rear silver metallization were deposited by magnetron sputtering. The front metallization was made of screen printed silver paste cured at 210 °C for 30 min. A total of five 2 × 2 cm$^2$ cells were made on each wafer. Additional details on the fabrication process can be found in [31].

After the cell metallization, the I–V characteristics were measured at STC (25 °C, 1000 W/m$^2$ AM1.5G spectrum) using a class AAA standard solar simulator. Additional I–V measurements from 15 to 80 °C with 5 °C steps were performed using LED and halogen-based solar simulator equipped with a thermoelectric holder and a PT100 temperature probe. For each of the 14 temperature steps, an I–V curve was measured at eight different light intensities, calibrated with a reference cell (the measurement conditions are shown in Fig. 4). The measurement uncertainty was probed by 25 successive measurements using a reference sample following the full measurement process each time (including removing and placing back the sample on the measurement setup), in the same conditions as the actual measurements presented in this article. For measurements performed on the same day (as was the case in this article), the incertitude of the temperature and lights intensity is measured to be well below 1 °C and 10 W/m$^2$, respectively. The standard deviation of the I–V parameters is the following: efficiency—0.033%$_{absolute}$, $V_{oc}$—0.075 mV, FF—0.078%$_{absolute}$, and $J_{sc}$—0.047 mA/cm$^2$.

The extensive measurement ranges enable to plot maps of the device efficiency as a function of temperature and irradiance (see Fig. 4). These maps were then used together with weather data to simulate the harvesting efficiency expected in various locations. The heat source $Q$ in the device was calculated as the fraction of the spectrally resolved incident absorbed irradiance (calculated using SunSolve from PVlighthouse [33]) minus the electrically produced power. Note that an iterative calculation was employed since PV conversion efficiency varies with temperature via the so-called TC. The device temperature was calculated using the ambient temperature $T_{amb}$, heat transfer coefficients ($u_0$, $u_1$) based on outdoor measurements taken from [34], and as a function of the wind velocity $v_w$ [2]

$$T_{module} = T_{amb} + \frac{Q}{u}. \quad (3)$$

With $u = u_0 + u_1 v_w$, the global heat transfer coefficient that accounts for all the heat transfer mechanisms. The produced power was calculated in 1-h-intervals over a complete year. The median of the operating conditions, weighted by the produced energy, is defined as the condition for which half of the harvested energy is produced at a higher temperature, half at lower temperature, half at higher illumination, and half at lower illumination [shown as white dots in Fig. 1(a)–(e)]. More details on the modeling framework can be found in [35]. Energy harvesting simulations were carried out in the locations, as shown in Fig. 1, representing different climatic conditions.

## III. RESULTS AND DISCUSSION

### A. Solar Cell Efficiency in Various Conditions

Fig. 3 shows I–V characteristics of the cells with different carbon content in the front doped layer, at STC and for the same illumination, but at a temperature of 60 °C. At STC, the cells with high-carbon content show a slightly larger $V_{oc}$ than the other two (i.e., carbon-free and low-carbon ones), yet this difference is quite small and could be linked to the fluctuation in the passivation quality. Then, the sample incorporating low (respectively, high) carbon concentration shows a 0.5 mA/cm$^2$ (respectively, 0.8 mA/cm$^2$) current gain compared with the reference cell, thanks to the more transparent a-SiC(n) layer. This higher transparency can also be measured by spectrometry of layers codeposited on glass (see Fig. 2) and originates from a wider bandgap [36]–[38]. However, the incorporation of carbon leads to an FF reduction through an increase of the $R_s$. This could possibly be caused by a larger conduction band offset or by a larger layer bulk resistivity [39]–[42]. At 25 °C, this higher resistance leads to a 2.5%$_{\text{absolute}}$ (respectively, 7.1%$_{\text{absolute}}$) FF loss compared with the reference sample. These competing effects lead to an efficiency 0.3%$_{\text{absolute}}$ (respectively, 1.4%$_{\text{absolute}}$) lower.

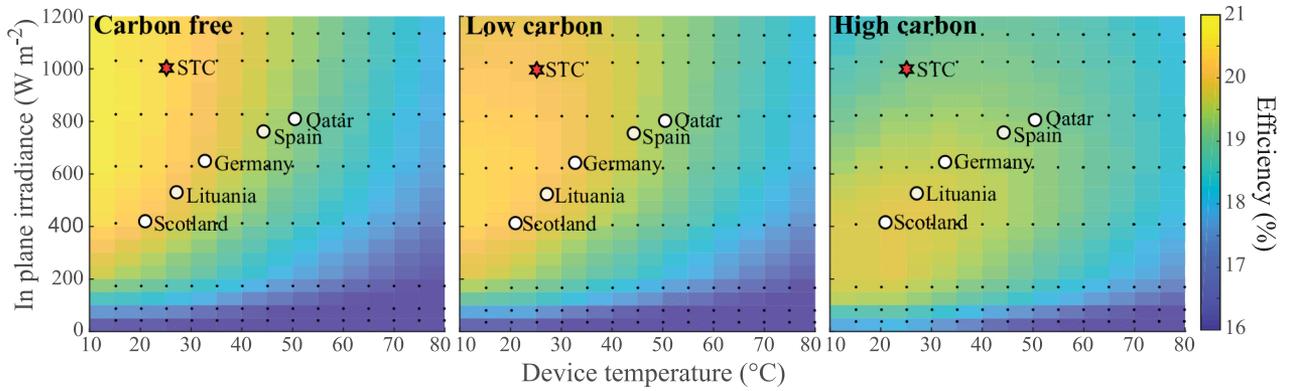

Fig. 4. Measured efficiency map as a function of the temperature and illumination for cells with different carbon concentrations in the front (n) layer. Each black dot represents one of the 9×14 measurement conditions for each sample.

When measuring the solar cell at 60 °C, $V_{\text{oc}}$ is lower and $J_{\text{sc}}$ has increased similarly for all samples, as would be expected from classical solar cell theory [43], thus not changing the trends between samples. Yet, the FF of all devices do not follow the $V_{\text{oc}}$ driven decrease as known from homojunction solar cells; either it does not change (for the carbon-free cell) or it even increases (for cells incorporating carbon). This stems from the presence of thermally activated resistances to charge extraction in these devices, as discussed in the introduction. As a result, the loss of the FF associated with carbon incorporation is much lower at 60 °C than 25 °C. The FF of the low-carbon (respectively, high-carbon) device is only 0.9%$_{\text{absolute}}$ (respectively, 2.9%$_{\text{absolute}}$) lower than the reference cell. This leads to an efficiency gain of 0.1%$_{\text{absolute}}$ (respectively, a loss of −0.1%$_{\text{absolute}}$) at 60 °C compared with the reference for the low-carbon (respectively, high-carbon) samples. We focus on the efficiency only and investigate mainly the effect of irradiances and temperatures of operation.

Fig. 4 shows the efficiency maps for the three devices measured at various operating temperatures and illumination intensities. The operating condition yielding the highest efficiency is shifting toward lower illumination and higher temperature upon carbon incorporation, as expected from the presence of aforementioned thermally activated transport barriers: Increasing temperature reduces the absolute value of $R_s$, whereas decreasing illumination decreases the impact of $R_s$ because of the increase of the impedance at the maximum power point [9]. Furthermore, since the current gain originating from the more transparent front layer stands at any temperature and illumination, the efficiency of the cells with carbon in the front (n) layer exceeds the one of the reference cells at high-enough temperature or low-enough illumination.

Fig. 5 further illustrates the difference between the various architectures: Fig. 5(a) shows an irradiance–temperature map of the subtraction between the efficiency of the carbon-free and the low-carbon samples, whereas Fig. 5(c)–(e) [respectively, Fig. 5(f)–(i)] shows efficiency values of the three samples at given temperatures (respectively, illuminations intensities). These views are indicative of the range for which a given carbon concentration in the (n) layer yields a better efficiency than the other one. For example, Fig. 5(f) shows that at 1000 W/m$^2$ (corresponding to 1 sun), the low-carbon cell efficiency exceeds the reference cell efficiency for temperatures over 50 °C, whereas at 400 W/m$^2$ [see Fig. 5(h)], the crossover occurs at 20 °C already. This lowering of the crossover temperature highlights that, especially in the case of heterojunction devices, the operating condition can change the hierarchy of the efficiency between devices employing different contact layers.

Back to Fig. 5(a), the upper-left corner of the measurement window (low temperature and high illumination) corresponds to conditions for which the carbon-free cell is more efficient as the low-carbon cell suffers from its higher $R_s$. However, this difference is reduced both by a lower illumination or a higher temperature, reaching a limit where the cells are equivalent [see white areas in Fig. 5(a)]. At these points, the resistive losses in the sample with carbon are compensated by the larger current. Reducing further the illumination intensity or increasing further the temperature, the low-carbon sample shows increasingly better efficiency than the reference cell. When comparing the carbon-free to the high-carbon cell [as shown in Fig. 5(b)], this trend is further amplified, shifting the equivalent-efficiency-line toward lower illumination and higher temperature. Thus, the efficiency gain at low illumination and high temperature is even larger than that recorded for the low-carbon cell. Note that there is not always a crossover point (or at least not for field-relevant conditions) since the addition of oxygen in the film in lieu of carbon yielded a systematic decrease of performance for all field-relevant temperature–illumination couples.

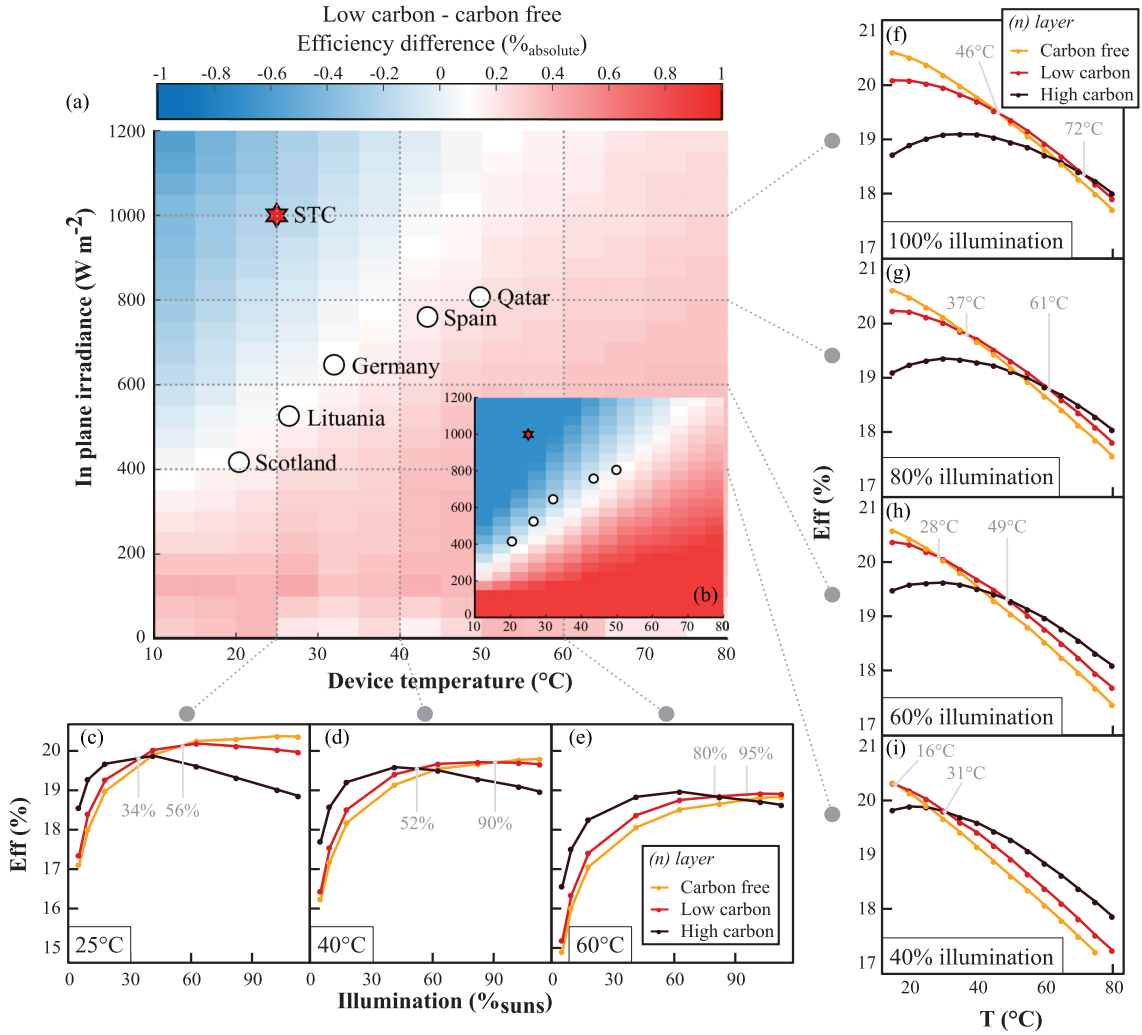

Fig. 5. (a), (b) Efficiency difference between the carbon-free cell and the cell with low-carbon content (n) layer (the inset in (b) shows the difference with the high-carbon content) as a function of the operating conditions. In the red zones, the low-carbon cell is more efficient than the carbon-free cell. The STC condition is highlighted with a red star, and the white dots are the energy-weighted median operating conditions of the different studied locations. (c)–(e) Efficiency as a function of the illumination for different temperatures. (f)–(i) Efficiency as a function of the temperature for different illumination.

### B. Harvesting Efficiency Calculations

In view of the different behavior of the efficiency with respect to the temperature and illumination for the various cell architectures, we calculated the total yearly average efficiency or harvesting efficiency [cf. (1)] output of the three investigated samples, in the climates mentioned above to allow a relevant comparison of their respective potential for electricity generation. These results are shown in Fig. 6: The sample with a low-carbon content, despite its lower efficiency at STC, produces more energy in all the considered climates. This harvesting efficiency difference is larger in the hot climates, reaching $+0.8\%_{\text{relative}}$ in Qatar. Interestingly, the low-carbon cell in moderate climates also shows a gain of about $+0.5\%_{\text{relative}}$, this time originating from the low illumination compared with STC. The sample with the high-carbon (n) layer is less efficient than the carbon-free sample in the moderate climates because of the resistive losses at low temperature. However, in very hot climates, such as Qatar, a gain is observed compared with the reference similar to the low-carbon samples. Thus, within the investigated conditions, the larger the carbon content is, the lower the STC efficiency is, but also the larger performance ratio is [see Fig. 6(c)] [44]. This overall results in similar harvesting efficiencies in outdoor conditions, confirming that the STC efficiency values can be misleading if the performance ratio is not considered.

Finally, it is worth noting that the yearly integrated harvesting efficiency differs from cell efficiency at the weighted median operating conditions. Using the median conditions for cell optimization is, thus, more realistic than considering solely STC conditions; however, the integration over the whole data is still needed for an accurate estimation of the harvesting efficiency. For a rapid estimation of an outdoor cell performance without computing the climatic data integration, an illumination of $700\,\text{W/m}^2$ and a temperature of $37\,°\text{C}$ (corresponding to the global mean operating conditions of all considered climates) can be used, as shown in Fig. 6.

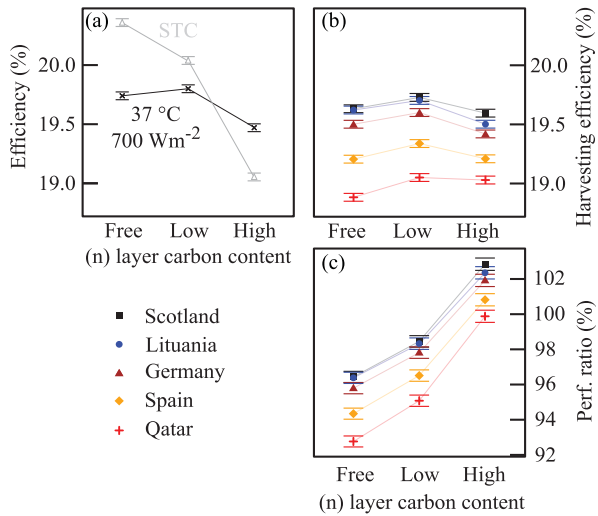

Fig. 6. (a) Efficiency at STC and at "realistic test conditions" (37 °C, 700 Wm$^{-2}$). (b) Harvesting efficiency as a function of the (n) layer carbon concentration in the different investigated climates. (c) Performance ratio as a function of the (n) layer carbon concentration in the different investigated climates. The error bars show the measurement uncertainty.

## IV. CONCLUSION

We compared the performance of silicon heterojunction solar cells using different carbon concentrations in the front (n) layer, leading to higher transparency but also higher resistive losses. We discuss the efficiency at different operating conditions and harvesting efficiency in various climates. We evidence that the mean operating conditions of a solar module in real outdoor situations do not match STC and are shifted to lower illumination for all the investigated climates, and to a usually higher temperature, particularly in hot climates. At STC, the resistive losses outweigh the optical gains, making the carbon-free cell the most efficient one. Upon increasing temperature or decreasing illumination, the optimum shifts toward the incorporation of carbon in the (n) layer. As a result, the device with a little carbon incorporation outperforms the reference device in all considered climates thanks to its larger harvesting efficiency. In moderate climates, such as Scotland, a 0.5%$_{relative}$ efficiency gain is observed which reached 0.8%$_{relative}$ in arid climates such as Qatar. These trends could not have been deduced from STC measurements, yet a measurement at a more realistic average operating condition at 37 °C and 700 W/m$^2$ (corresponding to the global mean operating conditions of all considered climates) would have hinted toward them. This highlights the need to calculate harvesting efficiencies in various climates when comparing solar cell designs.

## ACKNOWLEDGMENT

The authors would like to thank N. Badel, P. Wyss, and C. Allebé for the high-quality wet processing and metallization. They also thank C. Bucher and A. Schafflützel for technical support.